\documentclass[prb,aps,twocolumn,showpacs]{revtex4-2} 
\usepackage{amsmath,amssymb,amsthm}
\usepackage{graphicx}
\usepackage{subfigure}
\usepackage{amsmath}
\usepackage{amsfonts,amssymb}
\usepackage{bm}
\usepackage{float}
\usepackage{color}
\usepackage{sidecap}
\allowdisplaybreaks
\usepackage{soul}
\setstcolor{red}
\usepackage[normalem]{ulem}
\usepackage{hyperref}
\hypersetup{colorlinks=true,breaklinks,urlcolor=blue,linkcolor=blue,citecolor=blue}

\begin{document}

\title{Topological phase and its effective tuning in a ladder lattice}

\author{Qi-Bo Zeng}
\email{zengqibo@cnu.edu.cn}
\affiliation{Department of Physics, Capital Normal University, Beijing 100048, China}

\begin{abstract}
We study a two-leg ladder model consists of a one-dimensional (1D) Su-Schrieffer-Heeger (SSH) lattice with staggered nearest-neighboring hopping amplitudes and a normal 1D tight-binding lattice with uniform hopping. By varying the strength of inter-leg coupling, we find that topologically nontrivial phase with zero-energy edge modes will emerge, even when the SSH leg is in the trivial regime. Compared with the single SSH model, the nontrivial region in the parameter space is significantly expanded in the ladder. The topological phase is characterized by quantized Berry phase, and the phase boundaries are determined analytically. We also analyze the distributions of topological zero modes in the ladder, and find that the nontrivial regime can be further divided into two regions, which are separated by a gap closing point in the energy spectrum and correspond to the cases with edge modes residing in different legs. These results indicate that the topological phase and edge modes can be effectively tuned through the manipulations in the trivial lattice. Our work unveils the emergence of nontrivial topology in the ladder lattices and provides a new platform for studying topological phases.  
\end{abstract}
\maketitle
\date{today}

\section{Introduction}
During the past few decades, topological phases have garnered a lot of attention and have become one of the most fascinating research fields in condensed matter physics~\cite{Hasan2010RMP,Qi2011RMP,Ando2015ARCM,Elliott2015RMP}. With the presence of appropriate symmetries, many exotic topological phases have been predicted and realized, such as topological insulators~\cite{Bernevig2006Sci}, Weyl/Dirac semimetals~\cite{Hasan2017ARCM,YanARCM2017,Armitage2018RMP}, and more recently, higher-order topological insulators~\cite{Sitte2012PRL,Zhang2013PRL,Benalcazar2017PRL}, and so on. Due to the bulk-boundary correspondence, a topologically nontrivial system can host localized bound states at the boundaries, which are characterized by the topological invariants determined by the bulk band structures. To investigate the properties of topological systems, various tight-binding models have been constructed, such as the one-dimensional (1D) Su-Schrieffer-Heeger (SSH) model for solitons in polymers~\cite{Su1979PRL,Heeger1988RMP}, the 1D Kitaev chain for topological superconductors~\cite{Kitaev2001}, the Haldane model for quantum Hall effect~\cite{Haldane1988PRL}, and the Kane-Mele model for quantum spin Hall effect~\cite{Kane2005PRL}, etc. By varying the parameters of these models, the topological phases can be tuned systematically. Taking the celebrated SSH model as an example, if we adjust the relative strengths of the alternating hopping amplitudes, the system will go through a phase transition from the trivial to nontrivial phase or vice versa. With the rapid developments of the experimental techniques, the topological features like Zak phase and Thouless pumping can be directly observed in cold-atom systems~\cite{Atala2013NatPhy,Lohse2016NatPhy,Nakajima2016NatPhy}. In addition, topological phases have also been realized in photonic crystals~\cite{Wang2009Nature}, classical acoustic metamaterials~\cite{Susstrunk2015Science}, reconfigurable
microwave circuits~\cite{Peterson2018Nature}, and even mechanical systems of granular particles~\cite{Chaunsali2017PRL}, which enables the possibility of tuning topological phases under realistic circumstances.

On the other hand, the studies on topological phases have also been extended to the ladder lattices~\cite{Liu2012PRB,Li2013NatCom,Sun2016PRA,Ogino2021PRB,Ogino2022PRB,Mondal2023PRB,Parida2024PRB,Downing2024NJP,Aghtouman2024SciRep,Elia2025PRB}. Since the ladder is quasi-one dimensional, the system would behave differently due to the change of dimensionality. These models are important for investigating topological materials with layered structures or those coupled to substrates via the proximity effect, through which the interaction between the two systems provides an effective means of tuning the topological phases and can even give rise to novel topological phenomena~\cite{Lu2007PRB,Shoman2015NatCom,Hsieh2016PRL,Cheng2019PRB,Zheng2018PRB}. Compared with the 1D lattice, the ladder geometry provides more freedoms for tuning the topological phases. Due to the couplings between the two legs of the ladder, the properties of the original 1D system, such as the band structures, might be modified significantly. For instance, the existence of flat bands in the Creutz ladders~\cite{Sun2017PRB,Kuno2020NJP,Orito2021PRB,Mukherjee2022PRB,Pelegri2024PRB}. However, most studies to date have mainly focused on the models with two legs being the same kind lattice, e.g., the SSH~\cite{Nersesyan2020PRB,Zhang2017PRA,Li2017PRB,Padavic2018PRB,Jangjan2020SciRep,Jangjan2022PRB,Padhan2024PRB,Zhou2025PRB} and Kitaev ladders~\cite{DeGottardi2011NJP,Wu2012PLA,Maiellaro2018EPJS,Shibata2019PRB,Nehra2020PRR,Xu2024PRB}, which are composed of two SSH or Kitaev chains, respectively. It will be interesting to ask if we couple a topological lattice with a normal lattice, which is trivial by itself, what will happen to the nontrivial phase and topological edge modes. Meanwhile, it will also be important to check how we can tune the topological phase by manipulating the trivial lattice.

To answer these questions, in this work, we study the topological phases in a two-leg ladder lattice, which consists of the SSH model with staggered nearest-neighboring hopping and a simple normal tight-binding lattice with uniform hopping. Due to the inter-leg coupling, it is reasonable to speculate that when the SSH lattice is topologically nontrivial, the whole ladder could also be nontrivial. Here we revealed an unexpected phenomenon: when the SSH lattice is in the trivial regime, nontrivial phase can still emerge as we turn on the inter-leg coupling in the ladder. Most interestingly, compared with the single SSH lattice, the nontrivial region is significantly expanded in this ladder model. The trivial and nontrivial phases are characterized by quantized Berry phases, and the phase boundaries are determined analytically. We further analyze the distributions of zero-energy edge modes by setting different boundary conditions in the two legs, and find that the nontrivial phase is further divided into two different regions, which are separated by a gap closing point in the eigenenergy spectrum and correspond to the situations with edge modes existing in the SSH and the normal lattice, respectively. Thus, by manipulating the trivial lattice coupled with the SSH model, we can effectively tune the topological phase and edge modes. Our work demonstrates the emergent nontrivial topology and its effective tuning by the trivial lattice in the ladder systems.

The rest of this paper is organized as follows. In Sec.~\ref{Sec2}, we will first introduce the ladder model and give the model Hamiltonian as well as the dispersion relations. In Sec.~\ref{Sec3}, we discuss the variation of band structures as we vary the system parameters. Then we will investigate the nontrivial phase and the behaviors of zero-energy edge modes, and finally present the phase diagram. The last section (Sec.~\ref{Sec4}) is dedicated to a brief summary.

\begin{figure}[t]
	\includegraphics[width=3.0in]{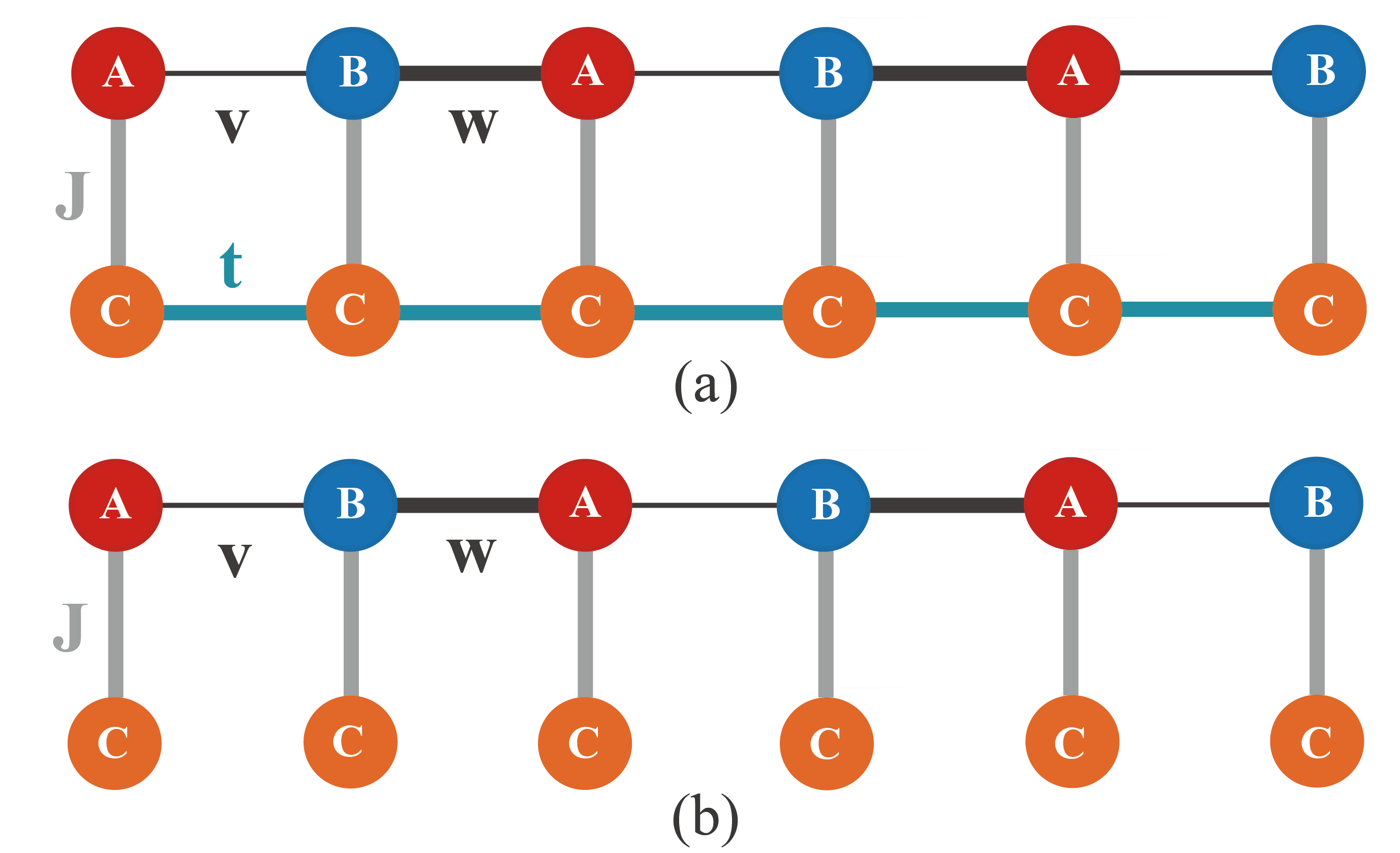}
	\caption{(Color online) (a) Schematic illustration of the ladder lattice studied in this work. The upper leg is the SSH lattice, with $v$ and $w$ being the staggered hopping between the $A$ and $B$ sites. The lower leg is a tight-binding lattice with uniform hopping between the $C$ sites. The inter-leg coupling is $J$, as denoted by the gray vertical lines. (b) A simplified model when the hopping between the $C$ sites in the lower leg is cut down by setting $t=0$, such that each site in the upper leg is coupled with a dangling site $C$.}
	\label{fig1}
\end{figure}

\section{Model Hamiltonian and energy spectrum}\label{Sec2}
We introduce a ladder model composed by the SSH lattice and a normal lattice. Fig.~\ref{fig1}(a) shows the schematic illustration of the ladder structure under open boundary conditions (OBC). The upper leg of the ladder is the SSH model contains $A$ and $B$ sites, with staggered modulation in the hopping amplitudes. The lower leg is a normal lattice with uniform hopping, where the site is labeled by $C$. The whole system is described by the following model Hamiltonian
\begin{equation}\label{H}
	H = H_{SSH} + H_{n} + H_{\perp},
\end{equation}
with
\begin{equation}
	\begin{aligned}
	H_{SSH} &= \sum_{j=1}^{N} \left(v \hat{a}_{j}^\dagger \hat{b}_{j} + w \hat{b}_{j}^\dagger \hat{a}_{j+1} + H.c. \right), \\ 
	H_{n} &= \sum_{j=1}^{N} \left( t \hat{c}_{2j-1}^\dagger \hat{c}_{2j} + t \hat{c}_{2j}^\dagger \hat{c}_{2j+1} + H.c. \right), \\
	H_{\perp} &= \sum_{j=1}^{N} \left(J \hat{a}_j^\dagger \hat{c}_{2j-1} + J \hat{b}_j^\dagger \hat{c}_{2j} + H.c. \right).
	\end{aligned}
\end{equation}
Here, $H_{SSH}$ describes the SSH lattice with $v$ and $w$ being the staggered hopping and $H_{n}$ describes the normal lattice with constant hopping amplitude $t$. $H_{\perp}$ gives the coupling between these two lattices, where the inter-leg coupling strength is given by $J$. $\hat{a}_{j}$, $\hat{b}_{j}$, and $\hat{c}_{j}$ ($\hat{a}_{j}^\dagger$, $\hat{b}_{j}^\dagger$, and $\hat{c}_{j}^\dagger$) are the annihilation (creation) operators of spinless fermions at the corresponding $A$, $B$ and $C$ site. Due to the coupling between the two legs, there are four sites in each unit cell. The lattice constant is set to $1$ and $N$ is the number of unit cells. The whole Hamiltonian in real space can be represented by a $L \times L$ dimensional matrix with $L=4N$. In this work, the parameters $v$, $w$, $t$, and $J$ are taken to be real numbers, and we will choose $w=1$ as the energy unit throughout this paper. 

It is well known that when the SSH model is topologically nontrivial, there will be zero-energy modes localized at the two ends of the lattice. Now that the SSH lattice is coupled with a normal lattice, we may expect the disappearance of these edge modes by tuning the system parameters, and thus making the whole system trivial. The unexpected fact we reveal in this paper is that, even when the SSH lattice in the ladder is in the trivial regime, the nontrivial phase still emerges as we turn on the inter-leg hopping and the uniform hopping in the lower leg. To better understand the emergence of nontrivial topology and zero-energy edge modes, we can start with a simplified model by cutting down the uniform hopping in the lower leg, i.e., setting $t=0$, as shown in Fig.~\ref{fig1}(b). Then we turn on the uniform hopping, and checking the variation of band structures. This will help us to get a more intuitive picture of the topological phase in the ladder lattice.

The energy spectra of the ladder under open boundary conditions (OBCs) is obtained by diagonalizing the Hamiltonian matrix directly. When the system is in the nontrivial phase, there will be in-gap edge modes localized at the ends of the ladder. To distinguish the edge modes from those extended bulk states, we can use the inverse participation ratio (IPR), which is defined as 
\begin{equation}
	\text{IPR}(\psi_m) = \sum_{j=1}^{L} |\psi_{m,j}|^4,
\end{equation}
where $\psi_m$ is the $m$th eigenstate with energy $E_m$, and $\psi_{m,j}$ is the $j$th component of the eigenstate. For a bulk state which is extended over the whole system, we have $\text{IPR} \rightarrow 0$, while for the topological edge states that are localized at the boundaries of the ladder, the IPRs are finite values of $O(1)$. Thus we can use the IPR to distinguish the edge states from the bulk states in the ladder.

Through Fourier transformation, we can transform the real space Hamiltonian shown in Eq.~(\ref{H}) into the momentum space and get the energy dispersion relations as well as the spectra under periodic boundary conditions (PBCs). Since there are four sites in each unit cell, there are four bands in the spectrum, and the Hamiltonian in the momentum space is a $4 \times 4$ matrix, which is
\begin{equation}\label{Hk}
	H_k = \begin{bmatrix}
		0 & r_1 & 0 & J \\
		r_1^* & 0 & J & 0 \\
		0 & J & 0 & r_2 \\
		J & 0 & r_2^* & 0
	\end{bmatrix},
\end{equation}
where we have set 
\begin{equation}\label{r12}
	r_1 = (v+w e^{-ik}), \qquad \text{and} \qquad r_2 = t(1+e^{ik}), 
\end{equation}
with $k \in [-\pi, \pi]$ being the Brillouin zone of the system. If we further introduce the Pauli matrices $\tau_i$ and $\sigma_i$ with $i=x,y,z$, which act in the ladder leg and the sublattice spaces of the SSH lattice, respectively, $H_k$ can be reformulated as
\begin{equation}
	\begin{aligned}
	H_k & = \frac{1}{2} \tau_z \otimes \left\{ \left[(v-t)+ (w-t) \cos k \right] \sigma_x + (w+t) \sin k \sigma_y  \right\} \\
		& + \frac{1}{2} I \otimes \left\{ \left[(v+t)+ (w+t) \cos k \right] \sigma_x + (w-t) \sin k \sigma_y  \right\} \\
		& + J \sigma_x \otimes \sigma_x.
	\end{aligned}
\end{equation}
Here $I$ is a $2 \times 2$ identity matrix. The Hamiltonian is chiral symmetric such that $S H_k S^{-1}=-H_k$, with $S=I \otimes \sigma_z$.

By diagonalizing the matrix $H_k$, we can obtain the four-band energy dispersion relation as follows
\begin{widetext}
\begin{equation}
	 E^2 = J^2 + \frac{1}{2} \left( |r_1|^2 + |r_2|^2 \right) \pm \frac{1}{2} \sqrt{\left( 2J^2 + |r_1|^2 + |r_2|^2 \right)^2 - 4 \left(J^2 - r_1 r_2 \right) \left(J^2 - r_1^* r_2^* \right)}.
\end{equation}
For the convenience of analysis, here we explicitly write down the four bands as
\begin{equation}\label{En}
	\begin{aligned}
		E_1 &= -\sqrt{J^2 + \frac{1}{2} \left( |r_1|^2 + |r_2|^2 \right) + \frac{1}{2} \sqrt{\left( 2J^2 + |r_1|^2 + |r_2|^2 \right)^2 - 4 \left(J^2 - r_1 r_2 \right) \left(J^2 - r_1^* r_2^* \right)}}, \\
		E_2 &= -\sqrt{J^2 + \frac{1}{2} \left( |r_1|^2 + |r_2|^2 \right) - \frac{1}{2} \sqrt{\left( 2J^2 + |r_1|^2 + |r_2|^2 \right)^2 - 4 \left(J^2 - r_1 r_2 \right) \left(J^2 - r_1^* r_2^* \right)}}, \\
		E_3 &= + \sqrt{J^2 + \frac{1}{2} \left( |r_1|^2 + |r_2|^2 \right) - \frac{1}{2} \sqrt{\left( 2J^2 + |r_1|^2 + |r_2|^2 \right)^2 - 4 \left(J^2 - r_1 r_2 \right) \left(J^2 - r_1^* r_2^* \right)}}, \\
		E_4 &= + \sqrt{J^2 + \frac{1}{2} \left( |r_1|^2 + |r_2|^2 \right) + \frac{1}{2} \sqrt{\left( 2J^2 + |r_1|^2 + |r_2|^2 \right)^2 - 4 \left(J^2 - r_1 r_2 \right) \left(J^2 - r_1^* r_2^* \right)}}, \\
	\end{aligned}
\end{equation}
where the four bands are labeled as $E_i$ with ($i=1,2,3,4$) from the lowest to the highest energy band. 
\end{widetext}

\section{Topological phase and edge modes in the ladder lattice}\label{Sec3}
In this section, we will start with the simplified models by cutting down the uniform hopping in the lower leg of the ladder, where the band structures and nonzero-energy edge modes will be investigated. Then we will turn on the hopping in the lower leg, and explore the emergent topological phase and the behaviors of the zero-energy edge modes in the ladder lattice.

\subsection{Properties of the simplified model}
\begin{figure}[t]
	\includegraphics[width=3.4in]{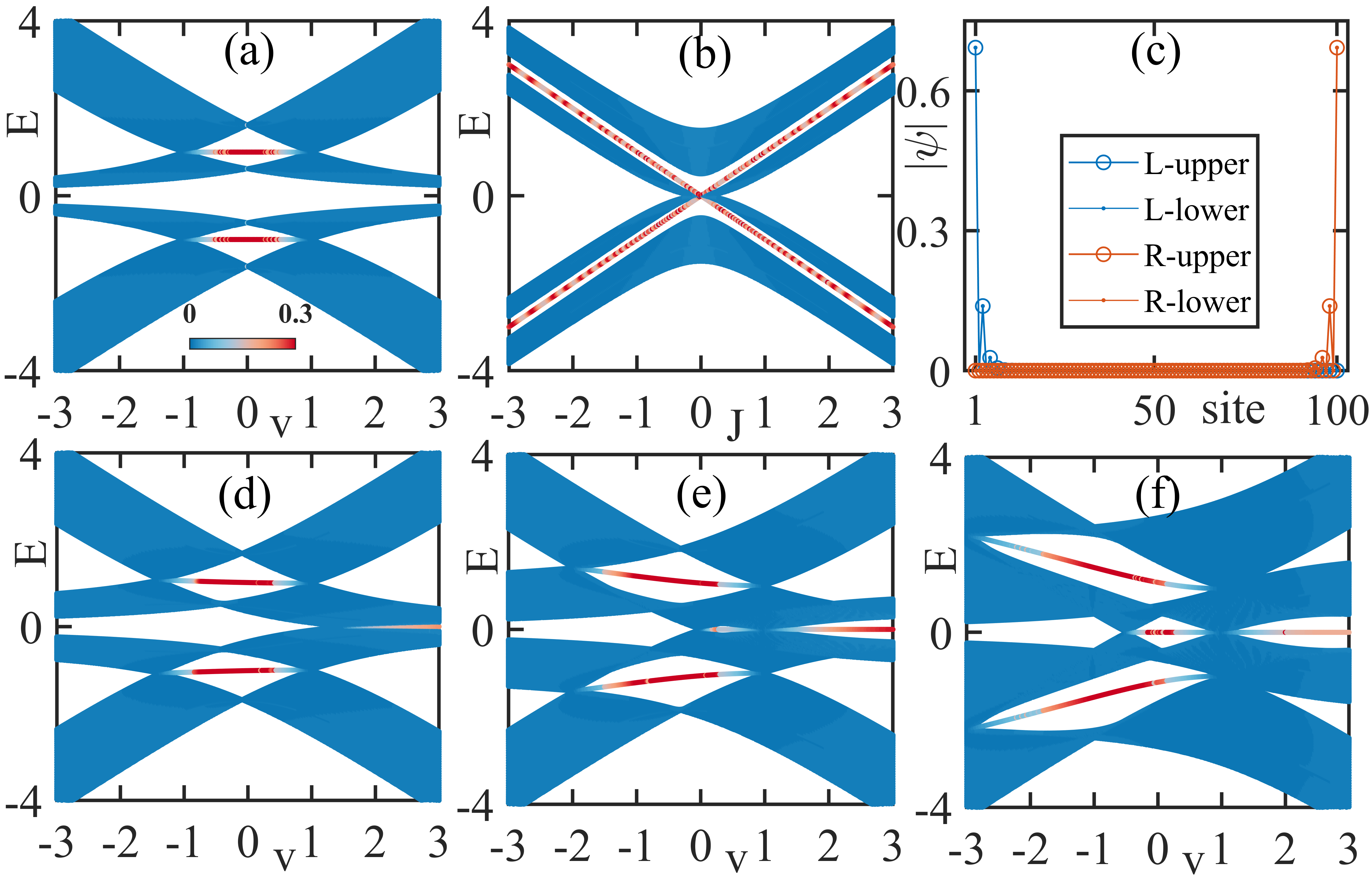}
	\caption{(Color online) Upper panel: OBC energy spectra and distribution of edge modes in the ladder model with $t=0$. (a) Eigenenergy as a function of $v$ when $J=1$; (b) Eigenenergy as a function of $J$ when $v=0.5$; (c) Distribution of the edge states shown in (a) at $v=0.2$. The empty blue circles (solid blue dots) represent the distributions at the left end of the upper SSH lattice (lower normal lattice), while the red ones denote the distributions at the right end. The colorbar indicates the IPR values of eigenstates. Lower panel: variations of the band structures for the ladder with $J=1$ as the uniform hopping $t$ increases: (d) $t=0.2$; (e) $t=0.5$; (f)$t=1.0$. Other parameters: $w=1$ and $N=50$.}
	\label{fig2}
\end{figure}

We first check the band structures of the simplified model under OBC shown in Fig.~\ref{fig1}(b), where we can set $t=0$ in the model Hamiltonian. In Fig.~\ref{fig2}(a), we plot the energy spectrum of the simplified model as a function of $v$ with the inter-hopping $J=1$. The colorbar in the figure denotes the IPR values (IPRs) of the eigenstates corresponding to the eigenenergies in the spectrum. 
 
From Fig.~\ref{fig2}(a), we can see that due to the coupling between the A/B sites in SSH lattice and the $C$ sites, the original two topological zero-energy modes in the SSH model within $|v|<|w|$ is split into four edge modes, where two with energy $E=J$ and the other two with $E=-J$. The distributions of these edge modes in the ladder are presented in Fig.~\ref{fig2}(c), where the edge modes resides on both legs with the same amplitudes. Note that under the condition $t=0$, the two bands near zero energy will never touch with each other, i.e., there will always be a gap between these two bands, which can be proved by using the formula for $E_2$ and $E_3$ in Eq.~(\ref{En}) with $t=0$. Figure \ref{fig2}(b) further presents the eigenenergies as a function of the inter-leg hopping strength $J$, and we can see that the edge modes vary linearly as $E=\pm J$, as indicated by the red dots. The linear variation of the edge modes can be explained as follows:  the single SSH lattice decoupled from the $C$ sites hosts topological zero-energy edge modes at the two ends of the lattice when $|v|<|w|$. Here we take the zero mode localized at the left end as an example, which is distributed on the $A$ sites and denoted as $| u_L \rangle = \sum_{j=1}^N a_j | j, A \rangle$ with $a_j$ being the probability amplitude such that $\sum_{j=1}^{N} |a_j|^2=1$. The state for the $C$ sites is set to be $|d_L \rangle = \sum_{j=1}^N \frac{1}{\sqrt{N}} | 2j-1, C \rangle$ with zero energy, which is uniform distributed before they are coupled with the upper leg. Next we turn on the couplings between the $A$ sites and the $C$ sites, and these two zero-energy states originally residing on $A$ and $C$ sites will hybridize. In the Hilbert space spanned by $| u_L \rangle$ and $| d_L \rangle$, we have 
\begin{equation}
	h = \begin{pmatrix}
	0 & \langle u_L | H_{int} | d_L \rangle \\ \langle d_L | H_{int} | u_L \rangle & 0  \end{pmatrix} 
	\propto \begin{pmatrix} 0 & J \\ J & 0 \end{pmatrix},
\end{equation} 
and the energies of the hybridized edge states become $E \propto \pm J$, consistent with the numeric results. Due to hybridization, the original degenerated zero-energy edge modes split into two nonzero-energy modes, and are evenly distributed on both $A$ and $C$ sites, as shown by the blue empty circles and blue solid dots in Fig.~\ref{fig2}(c). Similar analysis can also be applied to the edge modes localized at the right end.

\subsection{Emergent nontrivial phase in the ladder lattice}
Now we set $t \neq 0$ and turn on the uniform nearest-neighboring hopping between the $C$ sites in the lower leg of ladder.  Figure \ref{fig2}(d)-\ref{fig2}(f) present the energy spectrum as a function of $v$ for the system with $J=1$. As $t$ increases, the two central band $E_2$ and $E_3$ will get touched at zero energy and become gapless at certain points. Moreover, in addition to the nonzero-energy in-gap states, which are reminiscent of those discussed in the previous subsection, we also find zero-energy edge modes in the spectrum. Using the dispersion relations in Eq.~(\ref{En}), we can determine the gap closing points at zero energy analytically. Without loss of generality, we will set $t>0$ in the following discussions. Setting $E_2=E_3$, we have
\begin{equation}
	(J^2-r_1 r_2)(J^2-r_1^* r_2^*) = 0,
\end{equation}
where $r_1$ and $r_2$ are defined in Eq.~(\ref{r12}). By taking $k=0$, we find that the gap will close at
\begin{equation}\label{vc1}
	v_{c1} = \frac{J^2}{2t} - w.
\end{equation}
When $t$ is small, there will be only this one gap closing point, see Fig.~\ref{fig2}(d). And we can observe zero-energy edge modes in the regime $v>v_{c1}$. However, when $t>J^2/2w$, another gap closing point will emerge at 
\begin{equation}\label{vc2}
	v_{c2} = w.
\end{equation}
And the region within $v_{c1}<v<v_{c2}$ will also become gapped, and the zero edge modes also exist in this region, as shown in Fig.~\ref{fig2}(e) and \ref{fig2}(f). So, the variation of the uniform hopping between the $C$ sites in the lower leg can change the band structures and the topological properties of the ladder system. Note that when $v>w$, we know that the SSH lattice is in the trivial phase and there should be no edge modes in the system. Interestingly, by coupling the SSH lattice with a normal tight-binding lattice, nontrivial topological phase with zero-energy edge modes emerges in the originally trivial regime.

\begin{figure}[t]
	\includegraphics[width=3.4in]{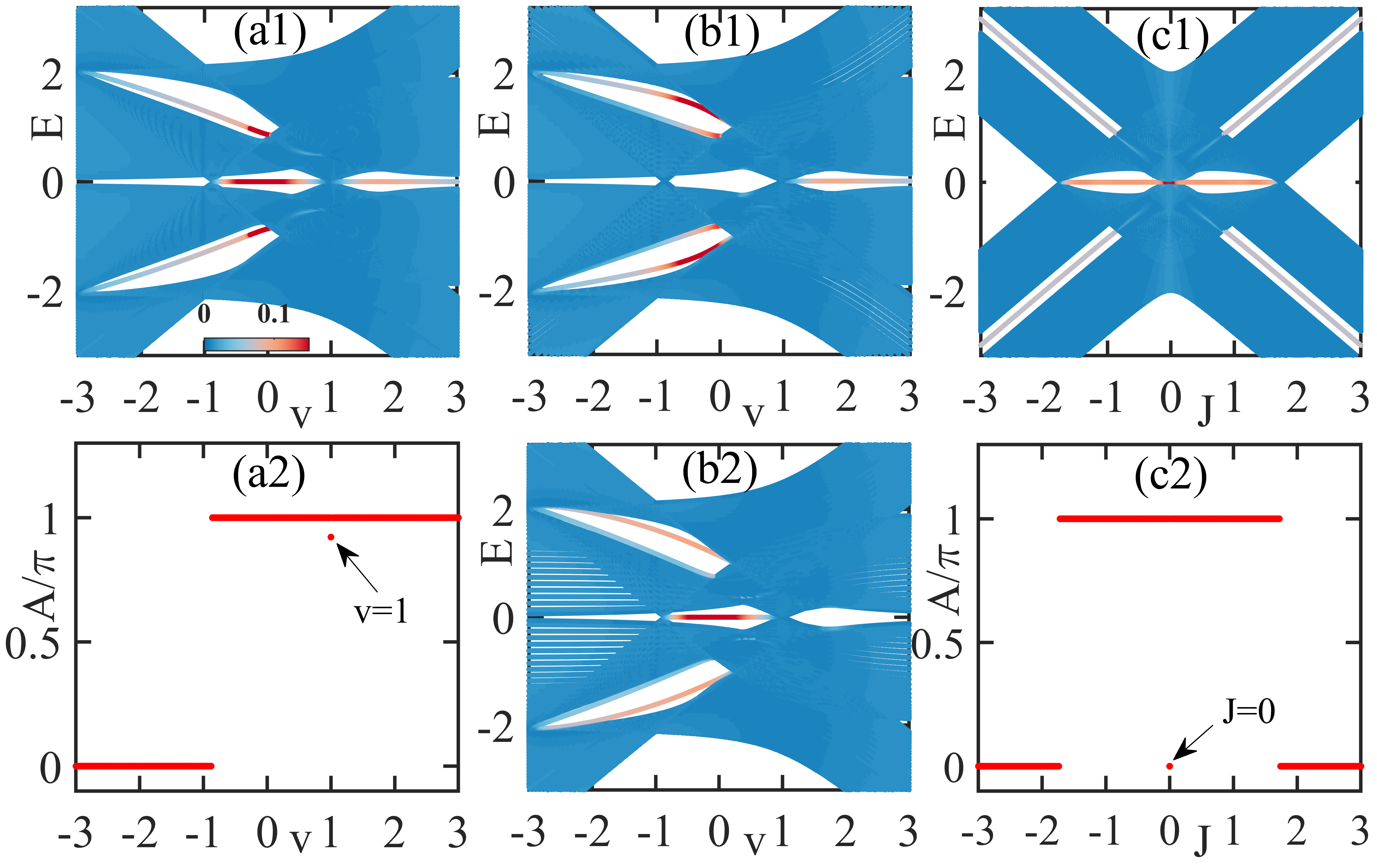}
	\caption{(Color online) (a1) The OBC energy spectrum as a function of $v$ for the ladder with $J=0.5$. The corresponding Berry phase is shown in (a2). The spectrum for the ladder in (a) with different boundary conditions: (b1) PBC in the upper SSH lattice and OBC in the lower normal lattice; (b2) OBC in the upper SSH lattice and PBC in the lower normal lattice. (c1) OBC spectrum as a function of $J$ for the ladder with $v=0.5$; (c2) the Berry phase. The arrows in (a2) and (c2) indicate the gap closing points, where the Berry phases are not well defined. Other parameters: $t=w=1$, and $N=50$.}
	\label{fig3}
\end{figure}

In Fig.~\ref{fig3}(a1), we plot the spectrum for the ladder with $J=0.5$ and $t=1$. Compared with the spectrum in Fig.~\ref{fig2}(f), we can find that as the inter-leg hopping $J$ gets stronger, the node at $v_{c1}$ will be shifted toward $v_{c2}$, and the topological region within these two nodes shrinks. Figure \ref{fig3}(c1) shows the spectrum as a function of $J$ when $v<w$. We can see that as the inter-leg coupling turns on, zero-energy edge modes will emerge in the region within $|J|<\sqrt{2t(v+w)}$. When $J$ further grows, the gap will close and the system becomes trivial. To characterize the topologically nontrivial phase with zero edge modes, we can calculate the Berry phase or Zak phase~\cite{Berry1984PRS,Zak1989PRL}, which is defined as
\begin{equation}
	A_s = \int_{-\pi}^{\pi} dk \langle \psi_s (k) | \partial_k \psi_s (k) \rangle, 
\end{equation} 
where $k$ is the momentum, $s=1,2,3,4$ is the band index, and $\psi_s(k)$ is the corresponding eigenvector of $H_k$. For the case we study here, we need to calculate the Berry phase for the lowest two bands with $E<0$, the Berry phase for the whole system is defined as $A=A_1 + A_2$, which characterizes the topological phase of the ladder system. Note that here $A$ is defined mod $2\pi$. The numeric results for the Berry phase $A$ as functions of $v$ and $J$ are presented in Fig.~\ref{fig3}(a2) and \ref{fig3}(c2). We can see that in the regime with (without) zero modes, we have $A=\pi$ ($0$). The sharp jump in the Berry phase indicates the topological phase transition from trivial to nontrivial phase, which are consistent with the gap closing points in the spectrum. In the region with $v>w$, as $v$ increases, the energy gap around zero energy will become smaller but always remain finite, thus the Berry phase always quantized at $\pi$. Note that the Berry phase at $v=w=1$ in Fig.~\ref{fig3}(a2) is not quantized since the gap closes at that point, which makes the Berry phase not well defined there. However, different from the critical point $v_{c1}$, the gap closing and reopening at $v_{c2}=w$ is not a topological phase transition since zero modes exist in the region $v_{c1}<v<v_{c2}$ and $v>v_{c2}$. Also the corresponding Berry phases are quantized at $\pi$ in both regions. However, the zero-energy edge modes in these two regions do behave differently, as we will discuss in the following.

In addition, in Fig.~\ref{fig3}(c1), we can also find nonzero-energy edge modes with the corresponding energies varying almost linearly with $J$. Note that if $v>w$, there will be no nonzero-energy edge modes in the spectrum (not shown here). As discussed above, these edge states are reminiscent of the nonzero-energy edge modes in the simplified model shown Fig.~\ref{fig1}(b), which is fully determined by the nontrivial region of the single SSH lattice. The existence of zero- and nonzero-energy edge modes have also been reported in the SSH chains with periodically modulated hoppings~\cite{Mandal2024PRB,Kar2024JPCM}.

\subsection{Behaviors of the zero-energy edge modes}
To demonstrate the different behaviors of zero-energy edge modes in the two topological regimes separated by $v_{c2} = w$, we first check the spectra of the ladder with different boundary conditions. Figure \ref{fig3}(b1) shows the spectrum for the ladder with the upper SSH lattice under PBC and the lower normal lattice under OBC. Comparing with the spectrum with both legs under OBCs [see Fig.~\ref{fig3}(a1)], we find that the zero modes in the region with $v_{c1}<v<v_{c2}$ disappear, while the zero modes in the region with $v>v_{c2}$ remain unchanged. On the other hand, if we set the upper leg with OBC but the lower leg with PBC, then the opposite situation happens. The zero modes in the $v>v_{c2}$ region disappear but those in the $v_{c1}<v<v_{c2}$ still exist. So, the nontrivial regime with $v_{c1}<v<v_{c2}$ is determined by the upper SSH lattice, while the nontrivial regime with $v>v_{c2}$ is determined by the lower leg, even though it is trivial itself. In addition, note that the nonzero-energy edge modes are also split in both cases, since they originate from the edge states in the simplified model shown in Fig.~\ref{fig1}(b), where the edge modes are evenly distributed in the $A$ (or $B$) and $C$ sites [see Fig.~\ref{fig2}(c)].  

\begin{figure}[t]
	\includegraphics[width=3.4in]{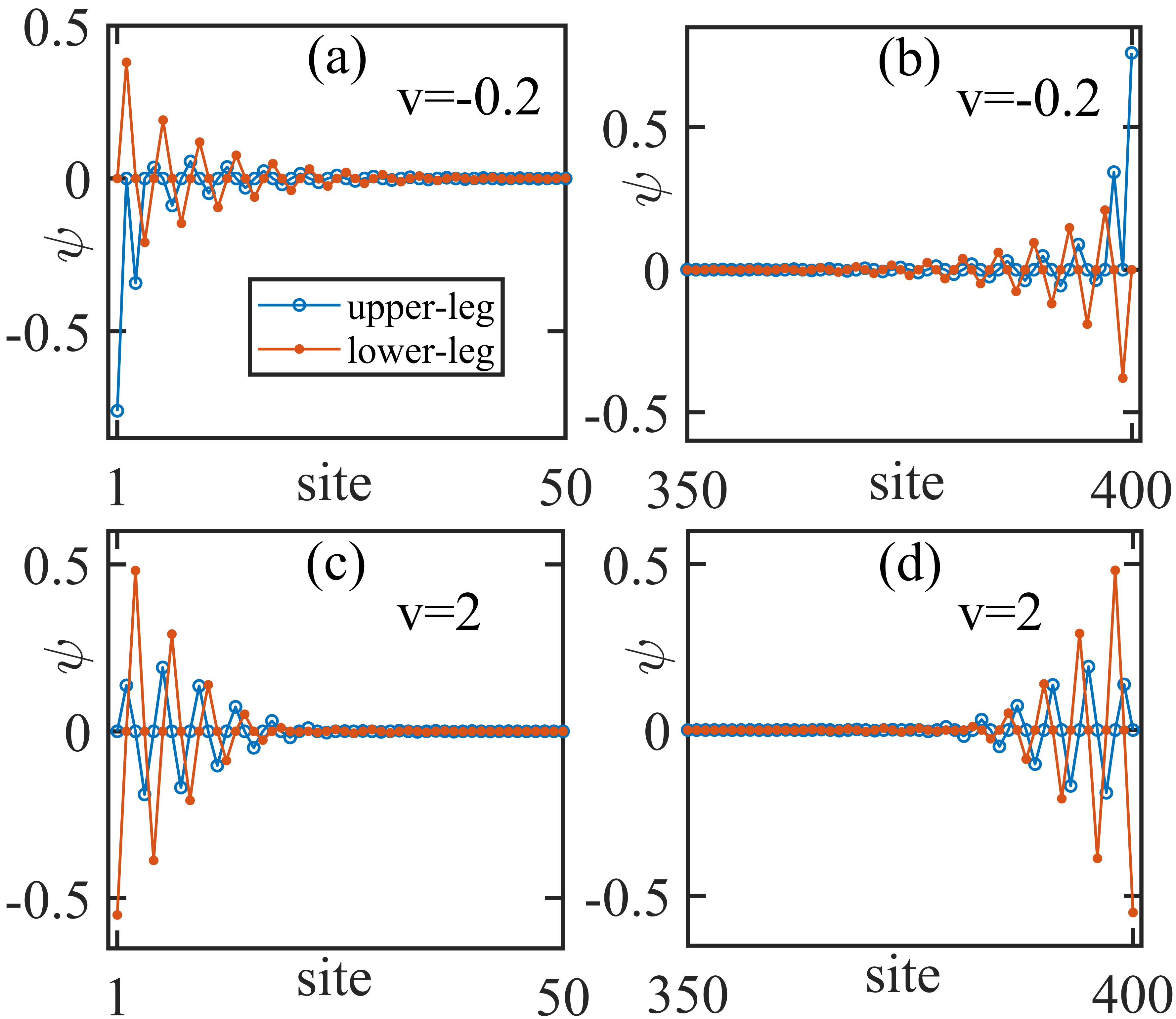}
	\caption{(Color online) Distributions of topological zero-energy edge modes in the two legs of the ladder lattice. Parameters are the same as those in Fig.~\ref{fig3}(a) but with a larger size $N=200$. Note that for clarity, we only present the region with the first or the last $50$ sites of each leg. (a) and (b) show that the edge modes exist in the left and right end of the upper leg, while in (c) and (d), the edge modes reside in the lower leg.}
	\label{fig4}
\end{figure}

We can further clarify the different behaviors of the edge modes in the two nontrivial regions by studying how the zero-energy edge modes distribute on the two legs of the ladder. Figure \ref{fig4} presents the numeric results for the system with the same parameters in Fig.~\ref{fig3}(a). Here we have chosen a larger system size with $N=400$. For the system with $v=-0.2$, which satisfies $v_{c1}<v<v_{c2}$, we find the two zero edge modes reside at the two ends of the upper SSH lattice, as represented by the blue empty circles in Fig.~\ref{fig4}(a) and \ref{fig4}(b). On the contrary, in the region $v>v_{c2}$, e.g., $v=2$, the edge modes exist in the lower leg, see the red solid circles in Fig.~\ref{fig4}(c) and \ref{fig4}(d). These behaviors are consistent with the spectrum properties we discuss above by setting different boundary conditions on the two legs. So, even though the critical point $v=w$ is not a topological phase transition point, it separates the topologically nontrivial region into two regimes, where the zero-energy edge modes reside in different legs of the ladder lattice.

\begin{figure}[t]
	\includegraphics[width=3.1in]{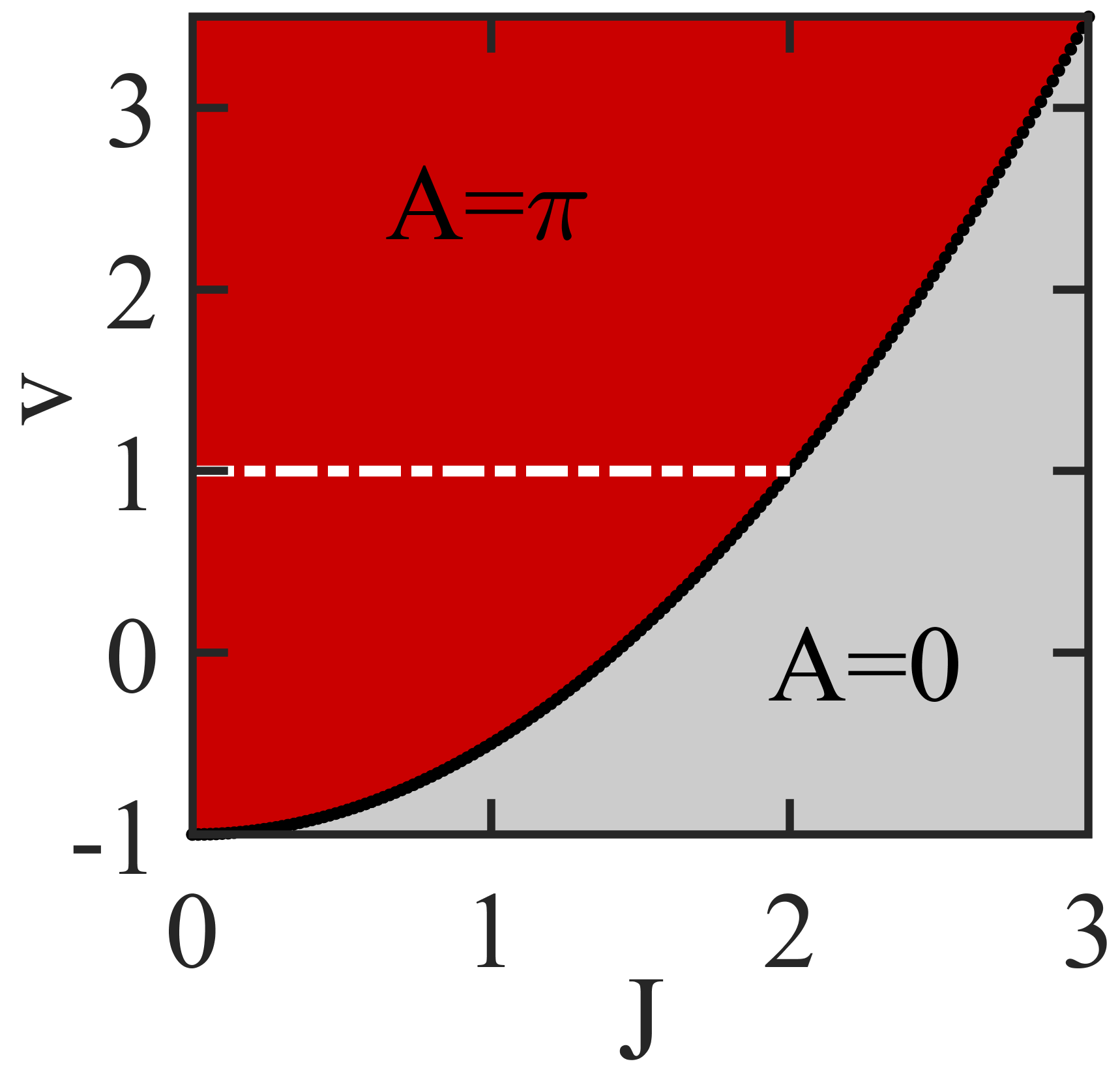}
	\caption{(Color online) Phase diagram of the ladder lattice with $w=t=1$. The red area is topologically nontrivial with Berry phase $A=\pi$, while the gray area is the trivial regime with $A=0$. The black line is the phase boundary that separates the trivial and nontrivial phase. The white dotted line corresponds to the gap closing point at $v=w$, which divided the nontrivial phase into two regions with the zero-energy edge modes existing in different legs.}
	\label{fig5}
\end{figure}

With all the discussions given above, here we give the phase diagram of the ladder lattice in the $v-J$ plane, see Fig.~\ref{fig5}. The red area is the topologically nontrivial region with Berry phase $A=\pi$, while the gray area is the trivial region with $A=0$. The trivial and nontrivial phase is separated by the critical line $v=\frac{J^2}{2t}-w$, as shown by the black line in the diagram. Note that the nontrivial region does not include the vertical line with $J=0$, because in this case, the two legs in the ladder are fully decoupled. So, by turning on the inter-leg coupling, we can always obtain topologically nontrivial phase if $v>\frac{J^2}{2t}-w$. The topological region is extensively expanded, compared with the single SSH chain with nontrivial region $v<w$. Besides, in the nontrivial phase, the white line corresponds to the critical point $v=w$, which separates the nontrivial phase with the zero edge modes existing in the upper or the lower leg of the ladder lattice. By tuning the inter-leg coupling and the uniform hopping in the trivial lattice, we can effectively tune the topological phase in the ladder model, where phases with different zero-energy edge modes can be obtained.

\section{Summary}\label{Sec4}
In summary, we have introduced a ladder model consists of the SSH lattice and a normal trivial lattice. After analyzing the energy spectra of the system with different parameters, we find that due to the inter-leg coupling between these two lattices, topologically nontrivial phase with zero-energy edge modes will emerge, even when the SSH lattice is trivial. The emergent nontrivial phase is characterized by nonzero quantized Berry phase. Moreover, we find that the nontrivial regime is divided into two regions by a gap closing point in the spectrum, which correspond to the cases with zero-energy edge modes residing in different legs. Thus, even though the energy gap closes and reopens at this critical point, it is not a topological phase transition. We have also discussed the properties of the nonzero-energy edge modes in the system. Finally, we presented the phase diagram of the ladder system. Our results demonstrate that by coupling the topological SSH model with a trivial normal lattice, we can effectively tune the topological phase and control the localization of edge modes in the two legs through the manipulations in the normal lattice, which make this ladder model more versatile in applications such as quantum control etc. Note that the ladder lattice introduced in this work can also be generalized to other situations by replacing the SSH model with other 1D topological systems, in which the topological phases can be tuned systematically. Our work unveils the emergent nontrivial phase in the ladder lattices and provides a new platform for studying topological phases or edge modes.

As to the experimental realizations, it is well known that the 1D SSH model has already been realized in various platforms such as cold-atom systems~\cite{Atala2013NatPhy,Lohse2016NatPhy,Nakajima2016NatPhy}, photonic crystals~\cite{Wang2009Nature}, acoustic memamaterials~\cite{Susstrunk2015Science}, and so on. It is feasible to realize our ladder model here by generalizing these systems, where the topological properties can be checked. Besides,  the topolectircal circuits, which are constructed by resistor, inductor and capacitor (RLC) components, have recently been extensively used to study various physical models~\cite{Lee2018ComPhy,Hofmann2019PRL,Zhang2025PRB,Jin2025PRL}. The circuits are easy to implement and to scale up, which makes them quite versatile in simulating physical models. Our ladder lattice here can also be simulated by using such circuits, where the circuit Laplacian plays the role model Hamiltonian. And the corresponding energy spectra and the topological edge modes under different boundary conditions could be detected and verified by measuring the impedance of the electrical circuits. 


\begin{acknowledgments}
This work is supported by the National Natural Science Foundation of China under Grant No. 12204326. 
\end{acknowledgments}


\begin{thebibliography}{}
\bibitem{Hasan2010RMP}{M. Z. Hasan and C. L. Kane, Colloquium: Topological insulators, \href{https://doi.org/10.1103/RevModPhys.82.3045}{Rev. Mod. Phys. \textbf{82,} 3045 (2010)}.}

\bibitem{Qi2011RMP}{X. L. Qi and S. C. Zhang, Topological insulators and superconductors, \href{https://doi.org/10.1103/RevModPhys.83.1057}{Rev. Mod. Phys. \textbf{83,} 1057 (2011)}.}

\bibitem{Ando2015ARCM}{Y. Ando and L. Fu, Topological crystalline insulators and topological superconductors: from concepts to materials, \href{https://doi.org/10.1146/annurev-conmatphys-031214-014501}{Annu. Rev. Condens. Matter Phys. \textbf{6,} 361 (2015)}.}

\bibitem{Elliott2015RMP}{S. R. Elliott and M. Franz, Colloquium: Majorana fermions in nuclear, particle, and solid-state physics, \href{https://doi.org/10.1103/RevModPhys.87.137}{Rev. Mod. Phys. \textbf{87,} 137 (2015)}.}

\bibitem{Bernevig2006Sci}{B. A. Bernevig, T. L. Hughes, and S. C. Zhang, Quantum spin Hall effect and topological phase transition in HgTe quantum wells, \href{https://www.science.org/doi/10.1126/science.1133734}{Science \textbf{314,} 1757 (2006)}.}

\bibitem{Hasan2017ARCM}{ M. Z. Hasan, S.-Y. Xu, I. Belopolsi, and S.-M. Huang, Discovery of Weyl fermion semimetals and topological Fermi arc states, \href{https://doi.org/10.1146/annurev-conmatphys-031016-025225}{Annu. Rev. Condens. Matter Phys. \textbf{8,} 289 (2017)}.}

\bibitem{YanARCM2017}{B. Yan and C. Felser, Topological materials: Weyl semimetals, \href{https://doi.org/10.1146/annurev-conmatphys-031016-025458}{Annu. Rev. Condens. Matter Phys. \textbf{8,} 337 (2017)}.}

\bibitem{Armitage2018RMP}{N. P. Armitage, E. J. Mele, and A. Vishwanath, Weyl and Dirac semimetals in three-dimensional solids, \href{https://doi.org/10.1103/RevModPhys.90.015001}{Rev. Mod. Phys. \textbf{90,} 015001 (2018)}.}

\bibitem{Sitte2012PRL}{M. Sitte, A. Rosch, E. Altman, and L. Fritz, Topological insulators in magnetic fields: quantum Hall effect and edge channels with a nonquantized $\theta$ term, \href{https://doi.org/10.1103/PhysRevLett.108.126807}{Phys. Rev. Lett. \textbf{108,} 126807 (2012)}.}

\bibitem{Zhang2013PRL}{F. Zhang, C. L. Kane, and E. J. Mele, Surface state magnetization and chiral edge states on topological insulators, \href{https://doi.org/10.1103/PhysRevLett.110.046404}{Phys. Rev. Lett. \textbf{110,} 046404 (2013)}.}

\bibitem{Benalcazar2017PRL}{W. A. Benalcazar, B. A. Bernevig, and T. L. Hughes, Quantized electric multipole insulators, \href{https://www.science.org/doi/10.1126/science.aah6442}{Science \textbf{357,} 61 (2017)}.}

\bibitem{Su1979PRL}{W. P. Su, J. R. Schrieffer, and A. J. Heeger, Solitons in polyacetylene, \href{https://doi.org/10.1103/PhysRevLett.42.1698}{Phys. Rev. Lett. \textbf{42,} 1698 (1979)}.}

\bibitem{Heeger1988RMP}{A. J. Heeger, S. Kivelson, J. R. Schrieffer, and W. P. Su, Solitons in conducting polymers, \href{https://doi.org/10.1103/RevModPhys.60.781}{Rev. Mod. Phys. \textbf{60,} 781 (1988)}.}

\bibitem{Kitaev2001}{A. Y. Kitaev, Unpaired Majorana fermions in quantum wires, \href{https://10.1070/1063-7869/44/10S/S29}{Phys. Usp. \textbf{44,} 131 (2001)}.}

\bibitem{Haldane1988PRL}{F. D. M. Haldane, Model for a quantum Hall effect without Landau levels: condensed-matter realization of the "parity anomaly", \href{https://doi.org/10.1103/PhysRevLett.61.2015}{Phys. Rev. Lett. \textbf{61,} 2015 (1988)}.}

\bibitem{Kane2005PRL}{C. L. Kane and E. J. Mele, Quantum spin Hall effect in graphene, \href{https://doi.org/10.1103/PhysRevLett.95.226801}{Phys. Rev. Lett. \textbf{95,} 226801 (2005)}.}

\bibitem{Atala2013NatPhy}{M. Atala, M. Aidelsburger, J. T. Barreiro, D. Abanin, T. Kitagawa, E. Demler, and I. Bloch, Direct measurement of the Zak phase in topological Bloch bands, \href{https://doi.org/10.1038/nphys2790}{Nature Phys. \textbf{9,} 795 (2013)}.}

\bibitem{Lohse2016NatPhy}{M. Lohse, C. Schweizer, O. Zilberberg, M. Aidelsburger, and I. Bloch, A Thouless quantum pump with ultracold bosonic atoms in an optical superlattice, \href{https://doi.org/10.1038/nphys3584}{Nature Phys. \textbf{12,} 350 (2016)}.}

\bibitem{Nakajima2016NatPhy}{S. Nakajima, T. Tomita, S. Taie, T. Ichinose, H. Ozawa, L. Wang, M. Troyer, and Y. Takahashi, Topological Thouless pumping of ultracold fermions, \href{https://doi.org/10.1038/nphys3622}{Nature Phys. \textbf{12,} 296 (2016)}.}

\bibitem{Wang2009Nature}{Z. Wang, Y. Chong, J. D. Joannopoulos, and M. Soljačić, Observation of unidirectional backscattering-immune topological electromagnetic states, \href{https://doi.org/10.1038/nature08293}{Nature (London) \textbf{461,} 772 (2009)}.}

\bibitem{Susstrunk2015Science}{R. Süsstrunk and S. D. Huber, Observation of phononic helical edge states in a mechanical topological insulator, \href{https://www.science.org/doi/10.1126/science.aab0239}{Science \textbf{349,} 47 (2015)}.}

\bibitem{Peterson2018Nature}{C. W. Peterson, W. A. Benalcazar, T. L. Hughes, and G. Bahl, A quantized microwave quadrupole insulator with topologically protected corner states, \href{https://doi.org/10.1038/nature25777}{Nature \textbf{555,} 346 (2018)}.}

\bibitem{Chaunsali2017PRL}{R. Chaunsali, E. Kim, A. Thakkar, P. G. Kevrekidis, and J. Yang, Demonstrating an in situ yopological band transition in cylindrical granular chains, \href{https://doi.org/10.1103/PhysRevLett.119.024301}{Phys. Rev. Lett. \textbf{119,} 024301 (2017)}.}

\bibitem{Liu2012PRB}{Z.-X. Liu, Z.-B. Yang, Y.-J. Han, W. Yi, and X.-G. Wen, Symmetry-protected topological phases in spin ladders with two-body interactions, \href{https://doi.org/10.1103/PhysRevB.86.195122}{Phys. Rev. B \textbf{86,} 195122 (2012)}.}

\bibitem{Li2013NatCom}{X. Li, E. Zhao, and W. Vincent Liu, Topological states in a ladder-like optical lattice containing ultracold atoms in higher orbital bands, \href{}{Nat. Commun. \textbf{4,} 1523 (2013)}.}

\bibitem{Sun2016PRA}{G. Sun, Topological phases of fermionic ladders with periodic magnetic fields, \href{https://doi.org/10.1103/PhysRevA.93.023608}{Phys. Rev. A \textbf{93,} 023608 (2016)}.}

\bibitem{Ogino2021PRB}{T. Ogino, S. Furukawa, R. Kaneko, S. Morita, and N. Kawashima, Symmetry protected topological phases and competing orders in a spin-1/2 XXZ ladder with a four-spin interaction, \href{https://doi.org/10.1103/PhysRevB.104.075135}{Phys. Rev. B \textbf{104,} 075135 (2021)}.}

\bibitem{Ogino2022PRB}{T. Ogino, R. Kaneko, S. Morita, and S. Furukawa, Ground-state phase diagram of a spin-1/2 frustrated XXZ ladder, \href{https://doi.org/10.1103/PhysRevB.106.155106}{Phys. Rev. B \textbf{106,} 155106 (2022)}.}

\bibitem{Mondal2023PRB}{S. Mondal, A. Agarwala, T. Mishra, and A. Prakash, Symmetry-enriched criticality in a coupled spin ladder, \href{https://doi.org/10.1103/PhysRevB.108.245135}{Phys. Rev. B \textbf{108,} 245135 (2023)}.}

\bibitem{Parida2024PRB}{R. Parida, A. Padhan, and T. Mishra, Interaction driven topological phase transitions of hardcore bosons on a two-leg ladder, \href{https://doi.org/10.1103/PhysRevB.110.165110}{Phys. Rev. B \textbf{110,} 165110 (2024)}.}

\bibitem{Downing2024NJP}{C A Downing, L Martín-Moreno and O I R Fox, Unconventional edge states in a two-leg ladder, \href{https://iopscience.iop.org/article/10.1088/1367-2630/ad5bf9}{New J. Phys. \textbf{26,} 073014 (2024)}.}

\bibitem{Aghtouman2024SciRep}{Sara Aghtouman and Mir Vahid Hosseini, Dimerized hofstadter model in two-leg ladder quasi-crystals, \href{https://doi.org/10.1038/s41598-024-59301-2}{Scientific Reports 14, 8782 (2024)}.}

\bibitem{Elia2025PRB}{João Pedro Gama D'Elia and Thereza Paiva, Topological phase transition in the two-leg Hubbard model: emergence of the Haldane phase via diagonal hopping and strong interactions, \href{https://doi.org/10.1103/rspn-4cyr}{Phys. Rev. B \textbf{112,} 035169 (2025)}.}

\bibitem{Lu2007PRB}{J. Lu and S. Wang, Tight-binding investigation of the metallic proximity effect of semiconductor metal double-wall carbon nanotubes, \href{https://doi.org/10.1103/PhysRevB.76.233103}{Phys. Rev. B \textbf{76,} 233103 (2007)}.}

\bibitem{Shoman2015NatCom}{T. Shoman, A. Takayama, T. Sato, S. Souma, T. Takahashi, T. Oguchi, K. Segawa, and Y. Ando, Topological proximity effect in a topological insulator hybrid, \href{https://doi.org/10.1038/ncomms7547}{Nat. Commun. \textbf{6,} 6547 (2015)}.}

\bibitem{Hsieh2016PRL}{T. H. Hsieh, H. Ishizuka, L. Balents, and T. L. Hughes, Bulk topological proximity effect, \href{https://doi.org/10.1103/PhysRevLett.116.086802}{Phys. Rev. Lett. \textbf{116,} 086802 (2016)}.}

\bibitem{Cheng2019PRB}{P. Cheng, P. W. Klein, K. Plekhanov, K. Sengstock, M. Aidelsburger, C. Weitenberg, and K. Le Hur, Topological proximity effects in a Haldane graphene bilayer system, \href{https://doi.org/10.1103/PhysRevB.100.081107}{Phys. Rev. B \textbf{100,} 081107(R) (2019)}.}

\bibitem{Zheng2018PRB}{Jun-Hui Zheng and Walter Hofstetter, Topological invariant for two-dimensional open systems, \href{https://doi.org/10.1103/PhysRevB.97.195434}{Phys. Rev. B \textbf{97,} 195434 (2018)}.}

\bibitem{Sun2017PRB}{N. Sun and L.-K. Lim, Quantum charge pumps with topological phases in a Creutz ladder, \href{https://doi.org/10.1103/PhysRevB.96.035139}{Phys. Rev. B \textbf{96,} 035139 (2017)}.}

\bibitem{Kuno2020NJP}{Y. Kuno, T. Orito, and I. Ichinose, Flat-band many-body localization and ergodicity breaking in the Creutz ladder, \href{https://doi.org/10.1088/1367-2630/ab6352}{New J. Phys. \textbf{22,} 013032 (2020)}.}

\bibitem{Orito2021PRB}{T. Orito, Y. Kuno, and I. Ichinose, Interplay and competition between disorder and flat band in an interacting Creutz ladder, \href{https://doi.org/10.1103/PhysRevB.104.094202}{Phys. Rev. B \textbf{104,} 094202 (2021)}.}

\bibitem{Mukherjee2022PRB}{A. Mukherjee, A. Nandy, S. Sil, and A. Chakrabarti, Tailoring flat bands and topological phases in a multistrand Creutz network, \href{https://doi.org/10.1103/PhysRevB.105.035428}{Phys. Rev. B \textbf{105,} 035428 (2022)}.}

\bibitem{Pelegri2024PRB}{G. Pelegrí, S. Flannigan, and A. J. Daley, Few-body bound topological and flat-band states in a Creutz ladder, \href{ https://doi.org/10.1103/PhysRevB.109.235412}{Phys. Rev. B \textbf{109,} 235412 (2024)}.}

\bibitem{Nersesyan2020PRB}{A. A. Nersesyan, Phase diagram of an interacting staggered Su-Schrieffer-Heeger two-chain ladder close to a quantum critical point, \href{https://doi.org/10.1103/PhysRevB.102.045108}{Phys. Rev. B \textbf{102,} 045108 (2020)}.}

\bibitem{Zhang2017PRA}{S.-L. Zhang and Q. Zhou, Two-leg Su-Schrieffer-Heeger chain with glide reflection symmetry, \href{https://doi.org/10.1103/PhysRevA.95.061601}{Phys. Rev. A \textbf{95,} 061601 (2017)}.}

\bibitem{Li2017PRB}{C. Li, S. Lin, G. Zhang, and Z. Song, Topological nodal points in two coupled Su-Schrieffer-Heeger chains, \href{https://doi.org/10.1103/PhysRevB.96.125418}{Phys. Rev. B \textbf{96,} 125418 (2017)}.}

\bibitem{Padavic2018PRB}{K. Padavić, S. S. Hegde, W. DeGottardi, and S. Vishveshwara, Topological phases, edge modes, and the Hofstadter butterfly in coupled Su-Schrieffer-Heeger systems, \href{https://doi.org/10.1103/PhysRevB.98.024205}{Phys. Rev. B \textbf{98,} 024205 (2018)}.}

\bibitem{Jangjan2020SciRep}{M. Jangjan and M. V. Hosseini, Floquet engineering of topological metal states and hybridization of edge states with bulk states in dimerized two-leg ladders, \href{https://doi.org/10.1038/s41598-020-71196-3}{Scientific Reports \textbf{10,} 14256 (2020)}.}

\bibitem{Jangjan2022PRB}{M. Jangjan and M. V. Hosseini, Topological properties of subsystem-symmetry-protected edge states in an extended quasi-one-dimensional dimerized lattice, \href{https://doi.org/10.1103/PhysRevB.106.205111}{Phys. Rev. B \textbf{106,} 205111 (2022)}.}

\bibitem{Padhan2024PRB}{A. Padhan, S. Mondal, S. Vishveshwara, and T. Mishra, Interacting bosons on a Su-Schrieffer-Heeger ladder: Topological phases and thouless pumping, \href{https://doi.org/10.1103/PhysRevB.109.085120}{Phys. Rev. B \textbf{109,} 085120 (2024)}.}

\bibitem{Zhou2025PRB}{Z. Zhou, Z.-C. Xu, and L.-J. Lang, Non-Abelian geometry, topology, and dynamics of a nonreciprocal Su-Schrieffer-Heeger ladder, \href{https://doi.org/10.1103/ykgv-y1y4}{Phys. Rev. B \textbf{112,} 094305 (2025)}.}

\bibitem{DeGottardi2011NJP}{W. DeGottardi, D. Sen and S. Vishveshwara, \href{https://iopscience.iop.org/article/10.1088/1367-2630/13/6/065028}{New J. Phys. \textbf{13,} 065028 (2011)}.}

\bibitem{Wu2012PLA}{N. Wu, Topological phases of the two-leg Kitaev ladder, \href{https://doi.org/10.1016/j.physleta.2012.10.016}{Phys. Lett. A \textbf{376,} 3530 (2012)}.}

\bibitem{Maiellaro2018EPJS}{A. Maiellaro, F. Romeo, and R. Citro, Topological phase diagram of a Kitaev ladder, \href{https://doi.org/10.1140/epjst/e2018-800090-y}{Eur. Phys. J. Spec. Top. \textbf{227,} 1397 (2018)}.}

\bibitem{Shibata2019PRB}{N. Shibata and H. Katsura, Dissipative spin chain as a non-Hermitian Kitaev ladder, \href{https://doi.org/10.1103/PhysRevB.99.174303}{Phys. Rev. B \textbf{99,} 174303 (2019)}.}

\bibitem{Nehra2020PRR}{R. Nehra, D. S. Bhakuni, A. Ramachandran, and A. Sharma, Flat bands and entanglement in the Kitaev ladder, \href{https://doi.org/10.1103/PhysRevResearch.2.013175}{Phys. Rev. Res. \textbf{2,} 013175 (2020)}.}

\bibitem{Xu2024PRB}{H. Xu and H.-Y. Kee, Reviving Majorana zero modes in the spin-1/2 Kitaev ladder model, \href{https://doi.org/10.1103/PhysRevB.110.195133}{Phys. Rev. B \textbf{110,} 195133 (2024)}.}

\bibitem{Berry1984PRS}{M. V. Berry, Quantal phase factors accompanying adiabatic changes, \href{https://doi.org/10.1098/rspa.1984.0023}{Proc. Roy. Soc. London A \textbf{392,} 45 (1984)}.}

\bibitem{Zak1989PRL}{J. Zak, Berry’s phase for energy bands in solids, \href{https://doi.org/10.1103/PhysRevLett.62.2747}{Phys. Rev. Lett. \textbf{62,} 2747 (1989)}.}

\bibitem{Mandal2024PRB}{S. Mandal and S. Kar, Topological solitons in a Su-Schrieffer-Heeger chain with periodic hopping modulation, domain wall, and disorder, \href{https://doi.org/10.1103/PhysRevB.109.195124}{Phys. Rev. B \textbf{109,} 195124 (2024).}}

\bibitem{Kar2024JPCM}{S. Kar, Edge state behavior in a Su–Schrieffer–Heeger like model with periodically modulated hopping, \href{https://doi.org/10.1088/1361-648X/ad0766}{J. Phys.: Condens. Matter \textbf{36,} 065301 (2024).}}

\bibitem{Lee2018ComPhy}{C. H. Lee, S. Imhof, C. Berger, F. Bayer, J. Brehm, L. W. Molenkamp, T. Kiessling, and R. Thomale, Topolectrical circuits, \href{https://doi.org/10.1038/s42005-018-0035-2}{Commun. Phys. \textbf{1,} 39 (2018)}.}

\bibitem{Hofmann2019PRL}{T. Hofmann, T. Helbig, C. H. Lee, M. Greiter, and R. Thomale, Chiral voltage propagation and calibration in a topolectrical Chern circuit, \href{https://doi.org/10.1103/PhysRevLett.122.247702}{Phys. Rev. Lett. \textbf{122,} 247702 (2019)}.}

\bibitem{Zhang2025PRB}{X. Zhang, C. Wu, M. Yan, and G. Chen, Observation of non-Hermitian pseudo-mobility-edge in a coupled electric circuit ladder, \href{https://doi.org/10.1103/PhysRevB.111.014304}{Phys. Rev. B \textbf{111,} 014304 (2025)}.}

\bibitem{Jin2025PRL}{W.-W Jin, J. Liu, X. Wang, Y.-R. Zhang, X. Huang, X. Wei, W. Ju, Z. Yang, T. Liu, and Franco Nori, Anderson delocalization in strongly coupled disordered non-Hermitian chains, \href{https://doi.org/10.1103/lpm2-vcb4}{Phys. Rev. Lett. \textbf{135,} 076602 (2025)}.}

\end{thebibliography}
\end{document}